\documentclass[11pt]{article} 

\usepackage{geometry} 
\usepackage{graphicx} 

\usepackage{booktabs} 
\usepackage{array} 
\usepackage{paralist} 
\usepackage{verbatim} 
\usepackage{subfig} 
\usepackage{amsmath}

\usepackage{fancyhdr} 
\usepackage{sectsty}
\allsectionsfont{\sffamily\mdseries\upshape} 

\usepackage{xcolor}

\title{Recombination velocities at grain boundaries in solar-cell absorbers - revisited}
\author{Daniel Abou-Ras$^1$ and Matthias Maiberg$^2$}
\date{%
    $^1$Helmholtz-Zentrum Berlin für Materialien und Energie GmbH, Germany\\%
    $^2$Martin-Luther-University Halle-Wittenberg, Germany\\[2ex]%
    \today
}

\begin{document}
\maketitle

\section*{Abstract}

The present work revisits the recombination velocities ($s_\mathrm{GB}$) of minority-charge carriers determined at grain boundaries in polycrystalline absorber materials for solar cells. The equations describing $s_\mathrm{GB}$ as well as the barriers for electrons and holes were derived. It is shown that for given net-doping density and absolute temperature, the experimentally determined recombination velocity of a specific grain boundary can be described by $s_{\mathrm{GB}} =  s_{\mathrm{GB,0}}^n \hspace{0.1cm} \mathrm{exp} \left[ - \Phi_{\mathrm{GB}}(N_{\mathrm{GB,charge}}) / (k_{\mathrm{B}} T) \right]$, where $\Phi_{\mathrm{GB}}$ is the band bending induced by the excess-charge density $N_{\mathrm{GB,charge}}$ at the grain boundary, and $k_{\mathrm{B}}$ as well as $T$ are the Boltzmann constant and the absolute temperature; i.e., $s_\mathrm{GB}$ depends only on the excess-charge density at this planar defect as well as on the prefactor $s_\mathrm{GB,0}^n$ describing the nonradiative recombination. Value ranges for these two quantities can be determined for any measured $s_\mathrm{GB}$ value. When analyzing $s_\mathrm{GB}$ datasets acquired on various (Ag,Cu)(In,Ga)Se$_2$ and microcrystalline Si absorbers, it is apparent that both, the excess-charge density and the prefactor $s_\mathrm{GB,0}^n$, remain within about the same orders of magnitude for all grain boundaries analyzed in a specific absorber. The broad range of the recombination velocities over several orders magnitude indicate upward as well as downward band bending, and the band-bending values are on the order of several $\pm$10 meV for all materials analyzed.  

\section{Introduction}

In the photovoltaic solar cells appropriate for low-cost productiion at high conversion efficiencies, most of the solar absorbers, with the exception of monocrystalline, wafer-based Si, are polycrystalline. Therefore, grain boundaries are present at densities corresponding to the average grain sizes in these absorbers.  Since the (two-dimensional, projected) density of point defects at a grain-boundary plane is enhanced with respect to those (three-dimensional) in the grain interiors of the adjacent grains, the nonradiative recombination rate is enhanced and contributes to a corresponding loss in the open-circuit voltage of the solar cell. The quantification of the grain-boundary recombination via the recombination velocity $s_\mathrm{GB}$ and the effects on the device performance have been revised recently \cite{quirk_2024}. It was clarified that $s_\mathrm{GB}$ can be expressed by a prefactor containing an effective defect density at the grain boundary and its capture cross-section multiplied by an exponential function depending on the band bending at the grain-boundary plane. However, for the case of a grain-boundary plane as opposed to for a surface, the derivation for equation describing the recombination velocity $s_\mathrm{GB}$ has not yet been published. Moreover, the available literature (e.g., \cite{seto_1975}) provides equations only for the barriers of the majority charge-carriers at the grain boundary (i.e., upward/downward band bending for $n$-type/$p$-type semiconductors).\\
\newline
The present work first derived the equations of $s_\mathrm{GB}$ for low-injection and high-injection conditions. Also, the barriers of the majority charge-carriers at the grain boundary were calculated. It became apparent that the barriers for electrons or holes, provided that the net-doping density is sufficiently high, depend for a given net-doping density and absolute temperature only on the excess-charge density at the grain boundary. It is highlighted that each $s_\mathrm{GB}$ value determined experimentally at a grain boundary can be simulated using appropriate values for the excess-charge density at the grain boundary and for the prefactor $ s_{\mathrm{GB,0}}^n$ describing the enhanced nonradiative recombination, and that the ranges of these values remain within the similar orders of magnitude for various photovoltaic absorber materials.

\section{Theory}

\subsection{Basic considerations}
\begin{figure}[h]
  \centering
  \includegraphics[width=8cm]{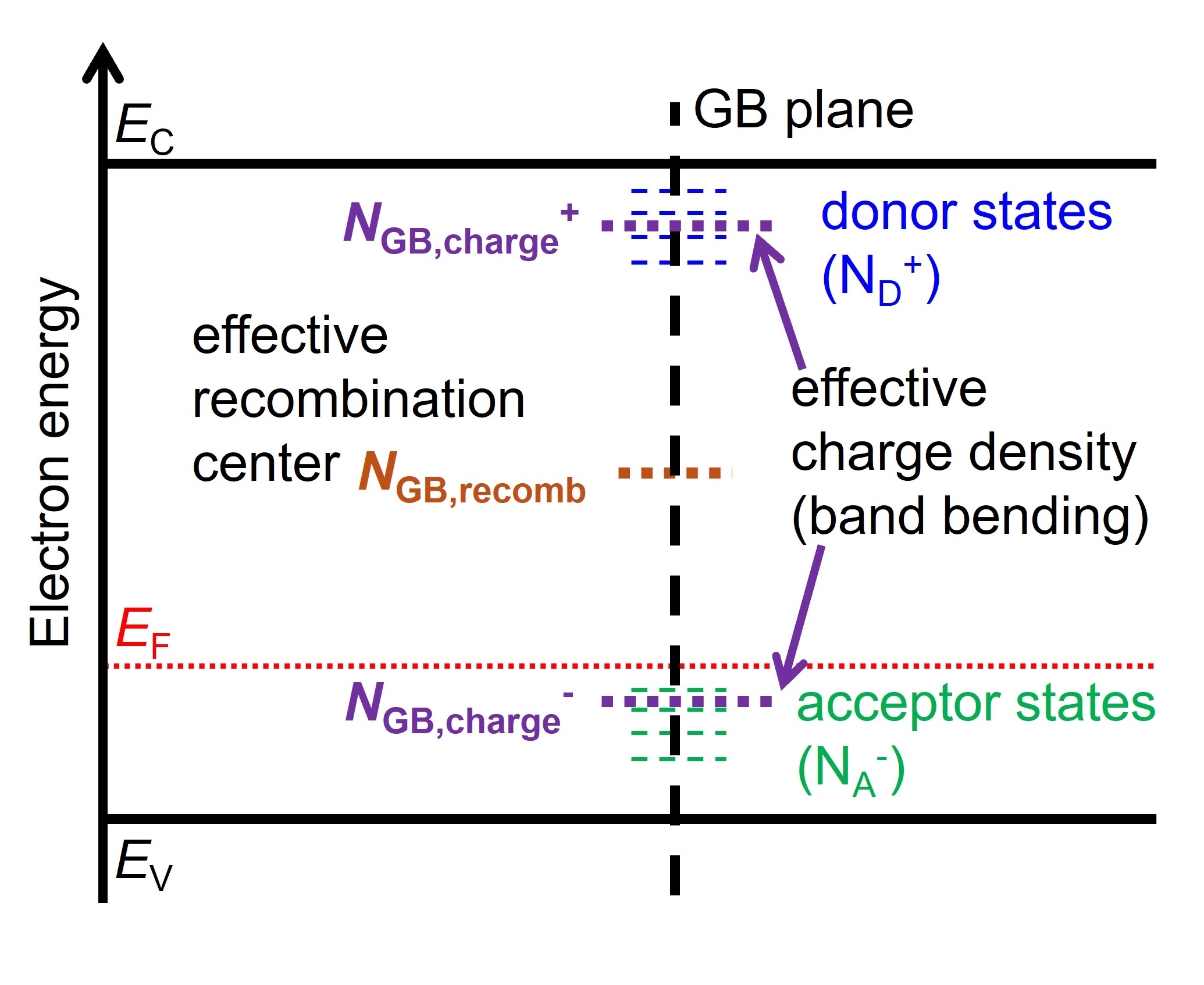}\\
\caption[]{Schematic equilibrium band diagram of a $p$-type semiconductor at room temperature around a grain-boundary (GB) plane. In case of a compensated semiconductor, positively charged donor ($N_{\mathrm{D}}^+$) as well as negatively charged acceptor states ($N_{\mathrm{A}}^-$) are present, in addition to neutral defect states (not depicted here). The GB model in the present work assumes a simplified scenario with one effective defect density $N_{\mathrm{GB,recomb}}$ at midgap position for the SRH recombination and another effective defect density $N_{\mathrm{GB,charge}}$, which can be positive or negative, depending on whether the density of positively or that of negatively charged defect states are larger. $E_{\mathrm{C}}$, $E_{\mathrm{V}}$, and $E_{\mathrm{F}}$ depict the conduction-band and valence-band edges as well as the Fermi level.}
 \label{fig:GB_band_diagram}
\end{figure}

We assume a $p$-type semiconductor with net-doping density $N_{\mathrm{A}}$.  This also includes the case of a compensated, $p$-type semiconductor, for which the doping density is $N_{\mathrm{a}} - N_{\mathrm{d}}$ (i.e., the difference between the densities of ionized acceptor and donor states). For this case, we write for the hole concentration in thermodynamic equilibrium:  $p_0 = N_{\mathrm{A}} = N_{\mathrm{a}} - N_{\mathrm{d}}$. Moreover, the electron density in the bulk under illumination is equal to the excess electron density: $n = n_0 + \Delta n \approx \Delta n$. The hole density in the bulk under illumination is equal to the sum of the net-doping density $N_{\mathrm{A}} $ and the excess electron density: $p = p_0 + \Delta p = N_{\mathrm{A}} +\Delta n$, provided that charge carriers are not separated by strong internal electrical fields..\\
It is also assumed that under illumination, the quasi-Fermi levels are constant throughout the semiconductor. Then, at a grain-boundary plane, the electron and hole densities can be expressed by
\begin{equation}
\label{eqn:gb_n}
n_{\mathrm{GB}} = \Delta n \hspace{0.1cm} \mathrm{exp} (-\Phi_{\mathrm{GB}}/k_{\mathrm{B}} T)
\end{equation}
and
\begin{equation}
\label{eqn:gb_p}
p_{\mathrm{GB}} = (N_{\mathrm{A}} +\Delta n) \hspace{0.1cm} \mathrm{exp} (\Phi_{\mathrm{GB}}/k_{\mathrm{B}} T),\\
\end{equation}
where $\Delta n$ is the electron density at the edges of the space-charge regions at the these planar defects, $\Phi_{\mathrm{GB}}$ is the band bending at the grain-boundary plane (which either drives electrons/holes to or repels them from the grain boundary), $k_{\mathrm{B}}$ is the Boltzmann constant, and $T$ the absolute temperature. 
\newline
We consider the following scenario at the grain-boundary. In the case of a compensated, $p$-type semiconductor at room temperature, there are various positively charged donor and negatively charged acceptor states, in addition to neutral defect states. Each defect state exhibits a charge state (positive, negative, or neutral) and contributes to Shockley-Read-Hall (SRH) recombination. The present grain-boundary model simplifies this complex situation by assuming one effective defect state at midgap position (since for such a defect, the SRH recombination is most effective) with density $N_{\mathrm{GB,recomb}}$ and its capture cross-sections for electrons and holes,  $\sigma_n$ and $\sigma_p$; and another effective defect state with density $N_{\mathrm{GB,charge}}$ beyond the demarcation levels. This effective defect state exhibits an excess charge corresponding to whether the density of positively or that of negatively charged defect states at the grain boundary are larger (see Fig. \ref{fig:GB_band_diagram}). 

\begin{figure}[h]
  \centering
  \includegraphics[width=15.5cm]{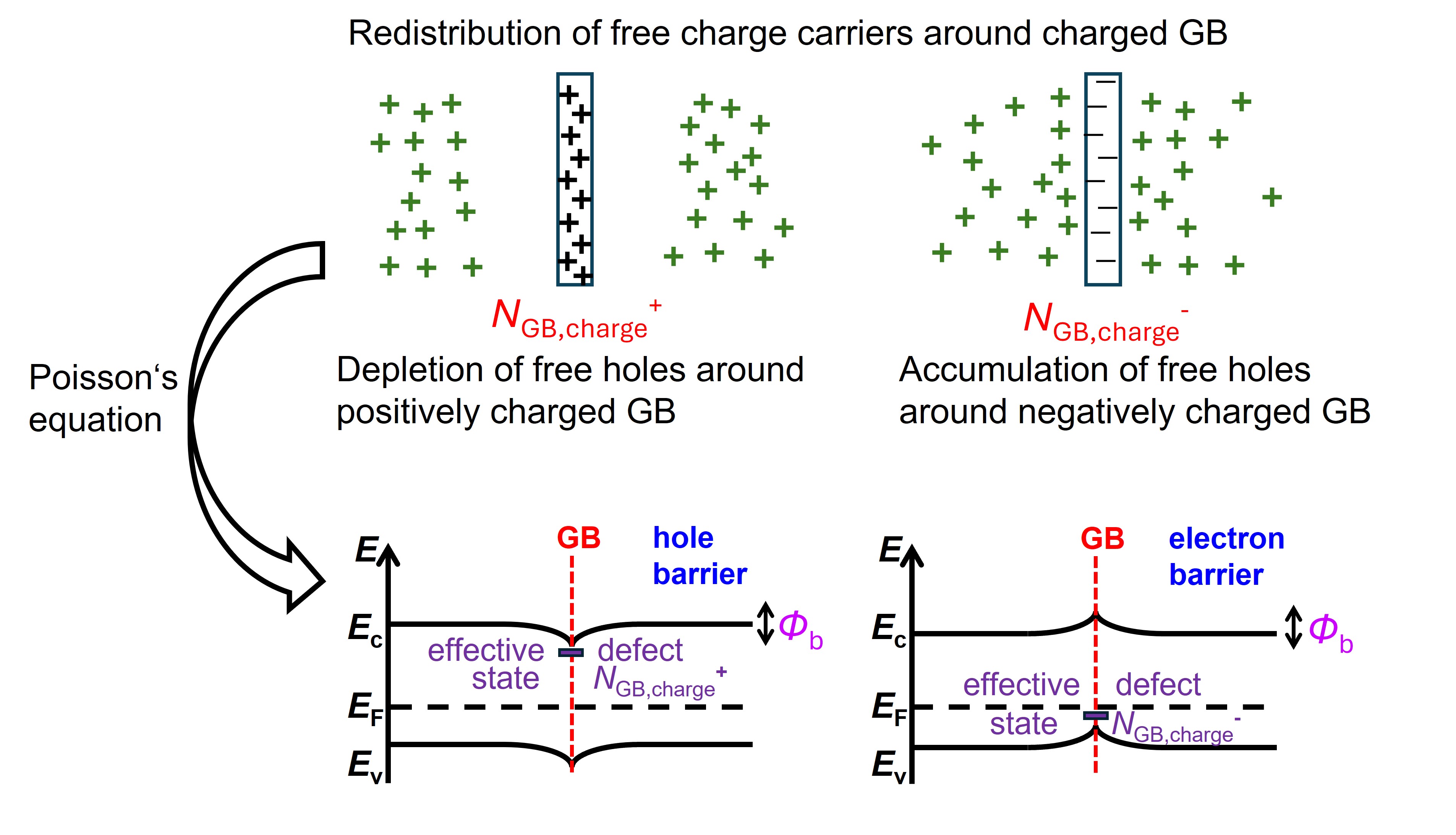}\\
\caption[]{Schematics depicting the origin of the band bending at a grain-boundary (GB) plane in a $p$-type semiconductor at room temperature. Free charge carriers (holes) redistribute correspondingly, provided that the net-doping density ($p_0$) is sufficiently large. According to Poisson's equation, the redistribution of free charge carriers results in downward (positive excess charge) or in upward band bending (negative excess charge). }
 \label{fig:GB_band-bending}
\end{figure}

Since the  effective, charged defect density $N_{\mathrm{GB,charge}}$ features a positive or negative excess charge, the free charge carriers redistribute correspondingly, provided that the net-doping density ($N_{\mathrm{A}}$) is sufficiently large (Fig. \ref {fig:GB_band-bending}). We note that for very small net-doping densities, the free charge carriers are captured by the defect state (trap), i.e., the grain interiors can be considered to be depleted of free charge carriers \cite{seto_1975}. According to Poisson's equation, the redistribution of free charge carriers results in spatial variations in the electrostatic potential and thus, in band bending. We note that for very large charge densities, theoretically, the upward band bending ($p$-type semiconductor) can take values above the Fermi level; however, the grain-boundary model in the present work does not consider such a scenario. 

\subsection{Derivation of the equation for the recombination rate at a grain boundary}

We use an equation describing the SRH recombination rate for an effective, deep defect level ($N_{\mathrm{GB,recomb}}$); it is very similar to the one used for the SRH bulk recomination rate $R_{\mathrm{SRH}} = \left(n p - n_i^2\right) / \left(n \tau_p + p \tau_n\right) $ ($\tau_p$ and $\tau_n$ are the hole and electron lifetimes, $n_i$ the intrinsic charge-carrier concentration). At a grain boundary, $n = n_{\mathrm{GB}}$ and $p = p_{\mathrm{GB}}$. Moreover, we assume $n_i^2 \ll n p$. The recombination rate at the grain boundary can be expressed by
\begin{equation}
\label{eqn:gb_recomb_1}
\begin{aligned}
R_{\mathrm{GB}} & = \frac{\Delta n \hspace{0.1cm} \mathrm{exp} (-\Phi_{\mathrm{GB}}/k_{\mathrm{B}} T) \hspace{0.1cm} (N_{\mathrm{A}} +\Delta n) \hspace{0.1cm} \mathrm{exp} (\Phi_{\mathrm{GB}}/k_{\mathrm{B}} T)}{\Delta n \hspace{0.1cm} \mathrm{exp} (-\Phi_{\mathrm{GB}}/k_{\mathrm{B}} T) \hspace{0.1cm} \left(s_{\mathrm{GB,0}}^{p}\right)^{-1} + (N_{\mathrm{A}} +\Delta n) \hspace{0.1cm} \mathrm{exp} (\Phi_{\mathrm{GB}}/k_{\mathrm{B}} T) \hspace{0.1cm} \left(s_{\mathrm{GB,0}}^{n}\right)^{-1}}\\ 
& = \frac{\Delta n (N_{\mathrm{A}} +\Delta n)}{\Delta n \hspace{0.1cm} \mathrm{exp} (-\Phi_{\mathrm{GB}}/k_{\mathrm{B}} T) \hspace{0.1cm}  \left(s_{\mathrm{GB,0}}^{p}\right)^{-1} + (N_{\mathrm{A}} +\Delta n) \hspace{0.1cm} \mathrm{exp} (\Phi_{\mathrm{GB}}/k_{\mathrm{B}} T) \hspace{0.1cm} \left(s_{\mathrm{GB,0}}^{n}\right)^{-1}}
\end{aligned}
\end{equation}
Here,  $s_{\mathrm{GB,0}}^{p}$ and $s_{\mathrm{GB,0}}^{n}$ are the recombination velocity parameters of holes and electrons at the grain boundary (we use expressions similar to those used at semiconductor surfaces \cite{aberle_2000}):
\begin{equation}
\label{eqn:S_p0}
s_{\mathrm{GB,0}}^{p} = N_{\mathrm{GB,recomb}} \hspace{0.1cm} \sigma_p \hspace{0.1cm} v_{\mathrm{th}}
\end{equation}
and
\begin{equation}
\label{eqn:S_n0}
s_{\mathrm{GB,0}}^{n} = N_{\mathrm{GB,recomb}} \hspace{0.1cm} \sigma_n \hspace{0.1cm} v_{\mathrm{th}},
\end{equation}
where  $v_{\mathrm{th}}$ is the thermal velocity of holes and of electrons (we assume that they are same). Using Eqs. \ref{eqn:S_p0} and \ref{eqn:S_n0}, Eq. \ref{eqn:gb_recomb_1} changes to:
\begin{equation}
\begin{aligned}
R_{\mathrm{GB}} & = \frac{\Delta n (N_{\mathrm{A}} +\Delta n)}{\Delta n \hspace{0.1cm} \mathrm{exp} (-\Phi_{\mathrm{GB}}/k_{\mathrm{B}} T) (N_{\mathrm{GB,recomb}} \sigma_p v_{\mathrm{th}})^{-1} + (N_{\mathrm{A}} +\Delta n) \hspace{0.1cm} \mathrm{exp} (\Phi_{\mathrm{GB}}/k_{\mathrm{B}} T) (N_{\mathrm{GB,recomb}} \sigma_n v_{\mathrm{th}})^{-1}}\\
& = \frac{\Delta n (N_{\mathrm{A}} +\Delta n)N_{\mathrm{GB,recomb}} v_{\mathrm{th}}}{\Delta n \hspace{0.1cm} \mathrm{exp} (-\Phi_{\mathrm{GB}}/k_{\mathrm{B}} T) \sigma_p^{-1} + (N_{\mathrm{A}} +\Delta n) \hspace{0.1cm} \mathrm{exp} (\Phi_{\mathrm{GB}}/k_{\mathrm{B}} T) \sigma_n^{-1}}\\
& = \frac{\Delta n N_{\mathrm{GB,recomb}} v_{\mathrm{th}}}{[\Delta n / (N_{\mathrm{A}} +\Delta n)] \hspace{0.1cm} \mathrm{exp} (-\Phi_{\mathrm{GB}}/k_{\mathrm{B}} T) \sigma_p^{-1} +  \mathrm{exp} (\Phi_{\mathrm{GB}}/k_{\mathrm{B}} T) \sigma_n^{-1}}
\end{aligned}
\end{equation}

\subsection{Equations for the recombination velocity at a grain boundary}
The following approach is corresponding to the work by Brody and Rohatgi \cite{brody_2001} who calculated the recombination velocity for a semiconductor surface. For a grain boundary, the effective recombination velocity is defined as
\begin{equation}
\label{eqn:sGB_def}
\begin{aligned}
s_{\mathrm{GB}} & := \frac{R_{\mathrm{GB}}}{\Delta n}\\
& = \frac{N_{\mathrm{GB,recomb}} v_{\mathrm{th}}}{[\Delta n / (N_{\mathrm{A}} +\Delta n)] \hspace{0.1cm} \mathrm{exp} (-\Phi_{\mathrm{GB}}/k_{\mathrm{B}} T) \sigma_p^{-1} +  \hspace{0.1cm} \mathrm{exp} (\Phi_{\mathrm{GB}}/k_{\mathrm{B}} T) \sigma_n^{-1}}.
\end{aligned}
\end{equation}
\newline
We may consider two different injection conditions:\\
\newline
\underline{Low-injection condition}\\ 
\newline
Here, we need to assume not only that $\Delta n \ll N_{\mathrm{A}}$, but also that the downward band bending $\Phi_{\mathrm{GB}}$ is not too strong; i.e., only if
\begin{equation}
\label{eqn:low_inj_cond}
\Delta n \ll N_{\mathrm{A}} / \left[ \mathrm{exp} (- 2\Phi_{\mathrm{GB}}/k_{\mathrm{B}} T) \sigma_n / \sigma_p - 1 \right] ,
\end{equation}
then $\Delta n / (N_{\mathrm{A}} +\Delta n)] \hspace{0.1cm} \mathrm{exp} (-\Phi_{\mathrm{GB}}/k_{\mathrm{B}} T) \sigma_p^{-1} \approx 0$ in Eq. \ref{eqn:sGB_def}, and this equation becomes
\begin{equation}
\label{eqn:s_GB}
s_{\mathrm{GB}} = N_{\mathrm{GB,recomb}} \hspace{0.1cm} \sigma_n v_{\mathrm{th}} \hspace{0.1cm} \mathrm{exp} \left( - \frac{\Phi_{\mathrm{GB}}}{k_{\mathrm{B}} T}\right).
\end{equation}
We note that in order for Eq.~\ref{eqn:low_inj_cond} to hold, $\Phi_{\mathrm{GB}}$ for the downward band bending must remain small. For $\Phi_{\mathrm{GB}}$ = -100 meV and $\sigma_n / \sigma_p \approx 1$, the factor $\left[ \mathrm{exp} (- 2\Phi_{\mathrm{GB}}/k_{\mathrm{B}} T) \sigma_n / \sigma_p - 1 \right]$ becomes about 3000. We consider $\Phi_{\mathrm{GB}}$ = -100 meV as a lower limit. 
\linebreak
\linebreak
\underline{High-injection condition}\\ 
\newline
$\Delta n \gg N_{\mathrm{A}}$: $\Delta n / (N_{\mathrm{A}} +\Delta n) \approx 1$\\
Also: $\Phi_{\mathrm{GB}} \approx 0$ since $n_{\mathrm{GB}} \approx p_{\mathrm{GB}}$\\

\begin{equation}
\label{eqn:s_GB_high_injection}
\Rightarrow s_{\mathrm{GB}} = N_{\mathrm{GB,recomb}} \hspace{0.1cm} v_{\mathrm{th}} (\sigma_p^{-1}+ \sigma_n^{-1})^{-1}.
\end{equation}

\subsection{The dependencies of $\Phi_{\mathrm{GB}}$ and $s_{\mathrm{GB}}$ vs. $N_{\mathrm{GB}}$ and  $\sigma_n$}

We note that for low-injection conditions, $s_{\mathrm{GB}}$ (Eq. \ref{eqn:s_GB}) is a function of $N_{\mathrm{GB,recomb}}$, of $\sigma_n$, and of $\Phi_{\mathrm{GB}}$. 
For downward band bending at the grain-boundary plane in a $p$-type semiconductor, $\Phi_{\mathrm{GB}}$ is expressed by \cite{seto_1975}
\begin{equation}
\label{eqn:GB_down}
\Phi_{\mathrm{GB}}^{\mathrm{down}} = - \frac{\left(q N_{\mathrm{GB,charge}}^+\right)^2}{8 \epsilon_0 \epsilon_r N_{\mathrm{A}}},
\end{equation}
where $\epsilon_0$ and $\epsilon_r$ are the dielectric permittivities of the vacuum and of the semiconductor (see also Sec. \ref{sec:app_downward_bb} in the Appendix). Note that Eq. \ref{eqn:GB_down} is only valid for semiconductors that exhibit net-doping densities sufficiently large so that the trap states represented by the density of recombination centers $N_{\mathrm{GB,recomb}}$ do not capture all free charge carriers; i.e., in case the grains in the polycrystalline semiconductor are only partly depleted.\\
The expression for the upward band bending is different (see Sec.  \ref{sec:app_upward_bb} in the Appendix):
\begin{equation}
\label{eqn:GB_up}
\Phi_{\mathrm{GB}}^{\mathrm{up}} = \frac{\sqrt{k_{\mathrm{B}}T} q N_{\mathrm{GB,charge}}^-}{4 \sqrt{\epsilon_0 \epsilon_{\mathrm{r}} N_{\mathrm{A}}}}.
\end{equation}
Note that for the derivation of \ref{eqn:GB_up}, the Boltzmann approximation was used; thus, $\Phi_{\mathrm{GB}}^{\mathrm{up}} < E_\mathrm{F} - E_\mathrm{V}$, which is about 100-150 meV for net-doping densities $N_{\mathrm{A}}$ of about $10^{14}$ to $10^{16}$ cm$^{-3}$. In the present work, we assume an upper limit for $\Phi_{\mathrm{GB}}^{\mathrm{up}}$ of 100 meV. \\
\newline
It is convenient to rewrite Eq. \ref{eqn:s_GB} for low-injection conditions using Eqs. \ref{eqn:S_p0} and \ref{eqn:S_n0}:
\begin{equation}
\label{eqn:s_GB_new}
\begin{aligned}
s_{\mathrm{GB}} =  s_{\mathrm{GB,0}}^n \hspace{0.1cm} \mathrm{exp} \left( - \frac{\Phi_{\mathrm{GB}}(N_{\mathrm{GB,charge}})}{k_{\mathrm{B}} T}\right).
\end{aligned}
\end{equation}
Thus, for given net-doping density $N_{\mathrm{A}}$ and absolute temperature $T$, $\Phi_{\mathrm{GB}}$ is a function of $N_{\mathrm{GB,charge}}$, and $s_{\mathrm{GB}}$ is a function of $N_{\mathrm{GB,charge}}$ and $s_{\mathrm{GB,0}}^n$. We can now vary both quantities within certain intervals and check the resulting value ranges for $\Phi_{\mathrm{GB}}$ and $s_{\mathrm{GB}}$. For the following considerations, we set $ \epsilon_r = 12$ and $T$ = 300 K. Corresponding to the considerations above, the current grain-boundary model is only valid for band bending values $\Phi_{\mathrm{GB}}$ within the interval of about -100 to +100 meV.\\

\begin{figure}[h!]
  \centering
  \includegraphics[width=12cm]{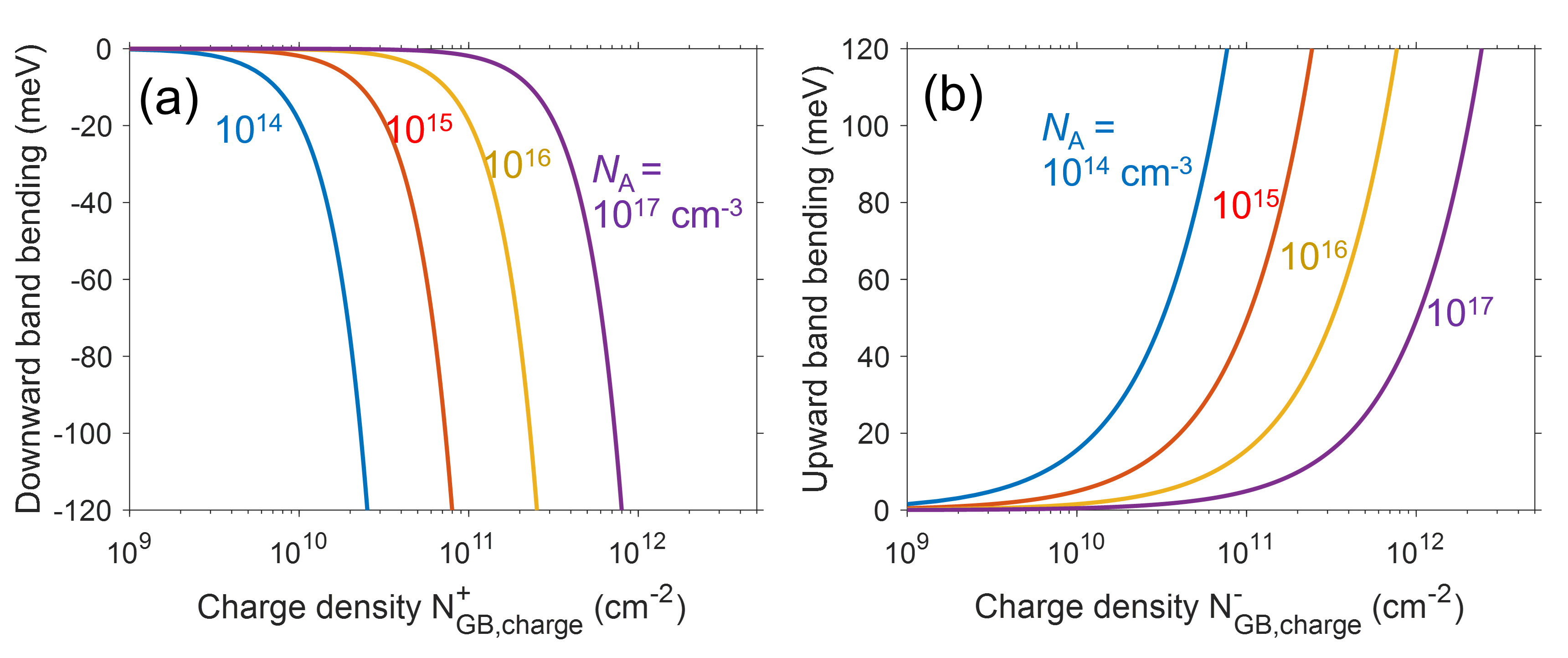}
\caption[]{(a) Downward band bending $\Phi_{\mathrm{GB}}^{\mathrm{down}}$ and (b) upward band bending $\Phi_{\mathrm{GB}}^{\mathrm{up}}$, calculated as a function of the charged defect density $N_{\mathrm{GB,charge}}$ for various net-doping densities $N_{\mathrm{A}}$ using Eqs. \ref{eqn:GB_down} and \ref{eqn:GB_up}. }
 \label{fig:phiB_NGB}
\end{figure}

While in microcrystalline Si, the $p$-type net-doping density is corresponding to the density of doping atoms (e.g., B for $p$-type Si),  in compensated semiconductors such as CdTe or CuInSe$_2$,  the $p$-type conductivity is determined by a slight excess of acceptor over donor densities, which are both on the order of about $10^{17}$-$10^{19} \hspace{0.1cm} \mathrm{cm}^{-3}$ \cite{neumann_1990,McCandless_2018,scarpulla_2023}. Assuming a grain-boundary width of $<$1 nm \cite{abou-ras_prl_2012}, the density of net charges at the grain-boundary plane would result to about $10^{9}$-$10^{11} \hspace{0.1cm} \mathrm{cm}^{-2}$. However, the general picture of a grain boundary is that it acts as a sink for point defects segregating during the growth from the grain interiors to these planar defects. Therefore, it is appropriate to assume a slightly larger range for $N_{\mathrm{GB,charge}}$ of about $10^{9}$-$10^{12} \hspace{0.1cm} \mathrm{cm}^{-2}$.  Since for our model to be valid, $\Phi_{\mathrm{GB}}^{\mathrm{down}}$ and $\Phi_{\mathrm{GB}}^{\mathrm{up}}$ must not exceed $\mp$ 100 meV, there is an upper limit for $N_{\mathrm{GB,charge}}$, which depends on the net-doping density $N_{\mathrm{A}}$. Using Eqs. \ref{eqn:GB_down} and \ref{eqn:GB_up}, the downward and upward band bending values $\Phi_{\mathrm{GB}}^{\mathrm{down}}$ and $\Phi_{\mathrm{GB}}^{\mathrm{up}}$ were calculated for various net-doping densities $N_{\mathrm{A}}$ (see Fig. \ref{fig:phiB_NGB}). Apparently, the upper limits for $N_{\mathrm{GB,charge}}$, i.e., for which the corresponding downward (upward) band bending values exceed $\mp$ 100 meV, are about $3 \times 10^{10}$ ($7 \times10^{10}$), $6 \times 10^{10}$ ($2 \times 10^{11}$), $3 \times10^{11}$ ($7 \times 10^{11}$), and $6 \times 10^{11}$ ($2 \times 10^{12}) \hspace{0.1cm} \mathrm{cm}^{-2}$ for $N_{\mathrm{A}} = 1 \times 10^{14}, 1 \times 10^{15}, 1 \times 10^{16}$, and $1 \times 10^{17} \hspace{0.1cm} \mathrm{cm}^{-3}$.


\begin{figure}[h!]
  \centering
  \includegraphics[width=14.6cm]{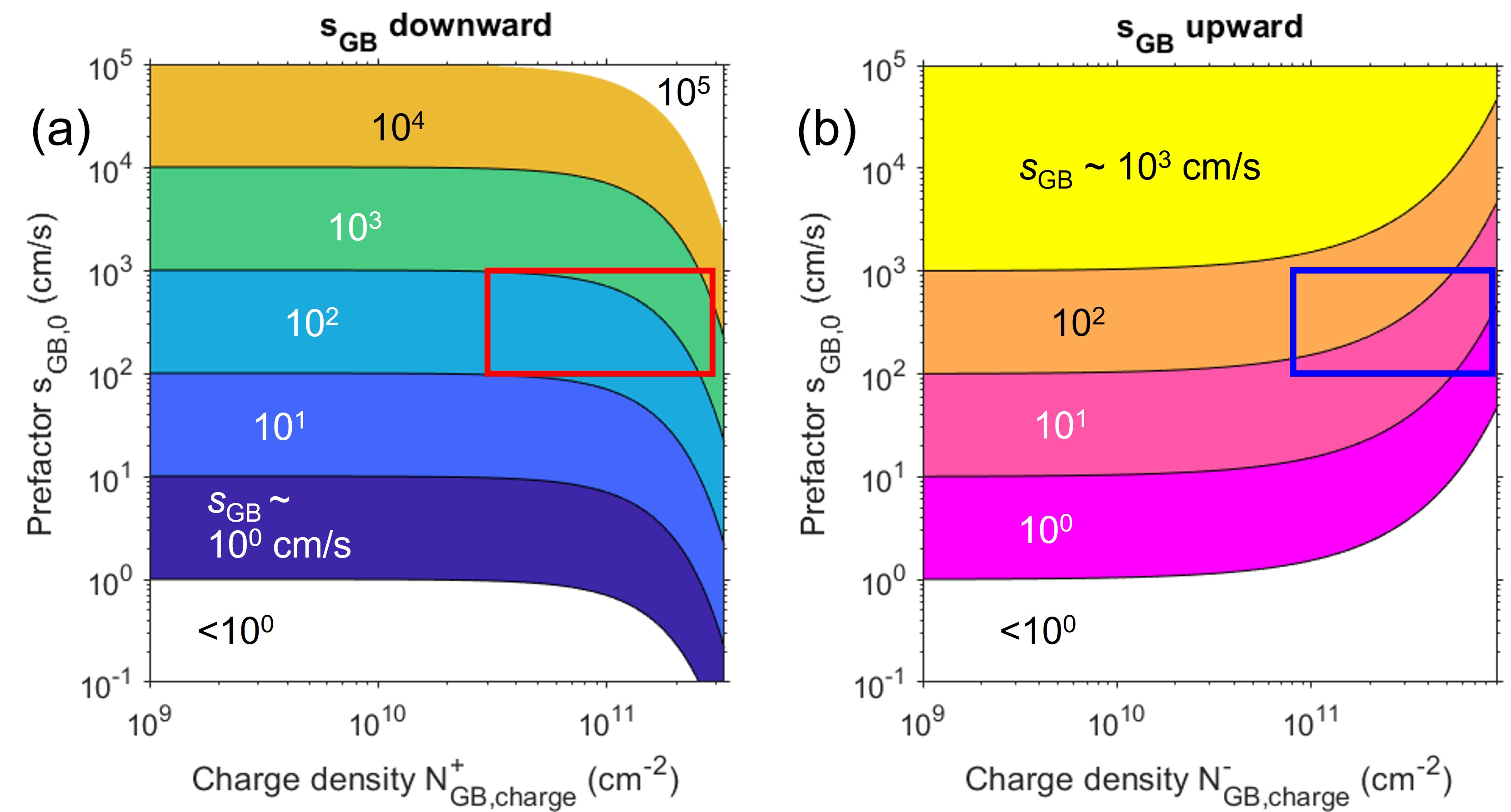}
\caption[]{The recombination velocity $s_{\mathrm{GB}}$ for (a) downward and (b) upward band bending as a function of the charged defect density $N_{\mathrm{GB,charge}}$ and of the prefactor $s_{\mathrm{GB,0}}^n$, calculated using Eqs. \ref{eqn:GB_down}, \ref{eqn:GB_up}, and \ref{eqn:s_GB_new} for $N_{\mathrm{A}} = 2 \times 10^{16} \hspace{0.1cm} \mathrm{cm}^{-3}$. Highlighted by red (a) and blue (b) rectangles are the value ranges for the evaluation of the experimental $s_{\mathrm{GB}}$ data given in Figure~\ref{fig:sGB_eval}.}
 \label{fig:sGB_sGB0_NGB}
\end{figure}

Assuming a net-doping density of $N_{\mathrm{A}} = 2 \times 10^{16} \hspace{0.1cm} \mathrm{cm}^{-3}$, $s_{\mathrm{GB}}$ for downward and upward band bending were calculated using Eq. \ref{eqn:s_GB_new}. Since the experimentally determined $s_{\mathrm{GB}}$ values that are on the order of $10^0$-$10^4$ cm/s \cite{sio_2016,krause_2020,thomas_2024_pip_1,thomas_2024_pip_2}, the appropriate value range for $s_{\mathrm{GB,0}}^n$ is about  $10^{-1}$-$10^{4} \hspace{0.1cm} \mathrm{cm/s}$. In Fig.~\ref{fig:sGB_sGB0_NGB}, we plotted the value range of $s_{\mathrm{GB}}$ as a function of the value ranges of $N_{\mathrm{GB,charge}}$ and $s_{\mathrm{GB,0}}^n$. We consider the upper limits for $N_{\mathrm{GB,charge}}$ with respect to the downward and upward band bending mentioned above, $3 \times10^{11}$ and $9 \times 10^{11} \hspace{0.1cm} \mathrm{cm}^{-2}$. The lower limit of $N_{\mathrm{GB,charge}}$ was set to $1 \times 10^{9} \hspace{0.1cm} \mathrm{cm}^{-2}$.

%

From Fig.~\ref{fig:sGB_sGB0_NGB}, it is clear that one specific $s_{\mathrm{GB}}$ value can be represented assuming a certain pair of $N_{\mathrm{GB,charge}}$ and $s_{\mathrm{GB,0}}^n$ values; and it is a range of appropriate values each that results in the same recombination velocity $s_{\mathrm{GB}}$.  
Table \ref{Tab:sGB_NGB_sigma} provides several ranges of $s_{\mathrm{GB}}$ values as well as the ranges of the corresponding $N_{\mathrm{GB,charge}}$ and $s_{\mathrm{GB,0}}$ values, assuming that they remain within the assumed intervals ($0.01$ to 3.3 or $0.01$ to $9.1 \times 10^{11} \hspace{0.1cm} \mathrm{cm}^{-2}$ and $1\times 10^{-1}$ to $1 \times 10^{5} \hspace{0.1cm} \mathrm{cm/s}$). 

\begin{table}[h!]
\caption{Ranges of the recombination velocity $s_{\mathrm{GB}}$ and the corresponding $N_{\mathrm{GB,charge}}$ and $s_{\mathrm{GB,0}}$ intervals, extracted from Figs. \ref{fig:sGB_sGB0_NGB}a and b. Note that the excess charge density represented by $N_{\mathrm{GB,charge}}$ is positive for downward and negative for upward band bending. }
\begin{center}
\begin{tabular}{c c c c} 
 \hline
Band-bending type & $s_{\mathrm{GB}}$ (cm/s) & $N_{\mathrm{GB,charge}} (10^{11} \hspace{0.1cm} \mathrm{cm}^{-2})$ & $s_{\mathrm{GB,0}}^n$ (cm/s)  \\ [0.5ex] 
 \hline
 Upward & 1-2 & 0.01-9.1 & 1-100\\ 
 \hline
Upward & 10-20 & 0.01-9.1 & 10-1000\\ 
 \hline
Upward & 100-200 & 0.01-9.1 & 100-10000 \\
 \hline
Downward & 100-200 & 0.01-3.3 & 2-200 \\
 \hline
Downward & 1000-2000 & 0.01-3.3 & 20-2000 \\
 \hline
Downward & 10000-20000 & 0.01-3.3 & 200-20000 \\ [1ex] 
 \hline
\end{tabular}
\label{Tab:sGB_NGB_sigma}
\end{center}
\end{table}

While the results depicted in Table \ref{Tab:sGB_NGB_sigma} are mathematically sound, the question is whether they can be confined to narrower intervals with respect to material properties in the solar-cell absorbers. In a polycrystalline semiconductor grown at elevated temperatures, the interdiffusion of constituing elements results in a rather homogeneous distribution of point defects, both, in the bulk and at grain boundaries. It does not mean identical, but certainly similar values for $N_{\mathrm{GB,charge}}$ and for $s_{\mathrm{GB,0}}^n$. Thus, it is appropriate to assume that within a specific absorber material, the $N_{\mathrm{GB,charge}}$ and $s_{\mathrm{GB,0}}^n$ values remain within the same orders of magnitude. In the following section, we will check whether under this assumption, experimental $s_{\mathrm{GB}}$ values from the same polycrystalline semiconductor can be successfully reproduced using Eq. \ref{eqn:s_GB_new}. 


\section{Comparison with experimental results}

The recombination velocity $s_{\mathrm{GB}}$ of a specific grain boundary in a polycrystalline semiconductor can be determined by evaluating cathodoluminescence (CL) intensity distributions across this planar defect. The corresponding procedure is described in Ref. \cite{quirk_2024}. 
As shown in the previous section, for a specific $s_{\mathrm{GB}}$ value, always, ranges of $N_{\mathrm{GB,charge}}$ and $s_{\mathrm{GB,0}}^n$ values can be determined. We will now apply this approach to experimentally measured $s_{\mathrm{GB}}$ values. (We note that in the evaluation procedure described in Ref. \cite{quirk_2024}, the median value of all experimental $s_{\mathrm{GB}}$ values from a specific absorber is assumed to be the $s_{\mathrm{GB}}$ value for $\Phi_{\mathrm{GB}}$ = 0 eV, i.e., equal to the prefactor $s_{\mathrm{GB,0}}^n$; however, we find that it is more appropriate to assume that  $s_{\mathrm{GB,0}}^n$ varies between two grain boundaries in the same material, and thus, this approach was improved by the present work.).\\

\begin{figure}[htbp!]
  \centering
  \includegraphics[width=14.5cm]{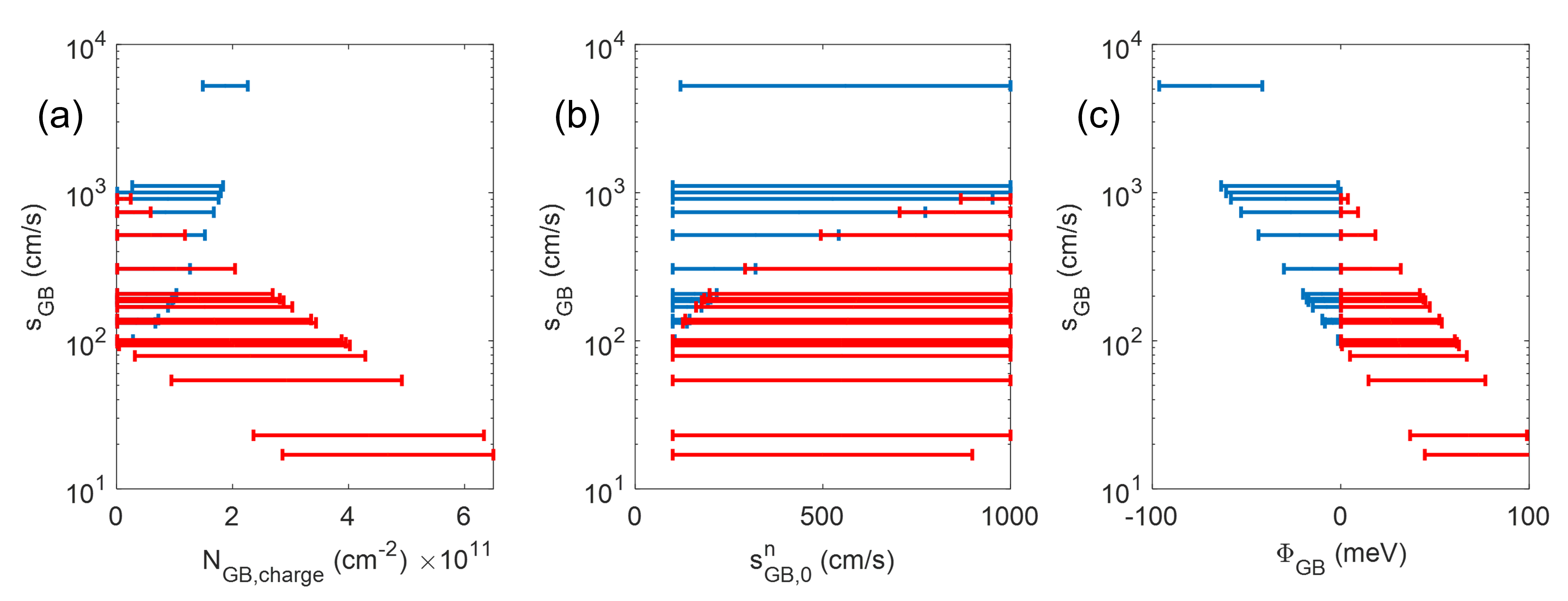}
\caption[]{Experimental $s_{\mathrm{GB}}$ values \cite{quirk_2024,krause_2020} as a function of (a) the defect density $N_{\mathrm{GB,charge}}$, of the prefactor $s_{\mathrm{GB,0}}^n$ (b), as well as of the band bending $\Phi_{\mathrm{GB}}$ (c). $N_{\mathrm{GB,charge}}$ was restricted to $0.3$-$3$ (downward) and $0.9$-$9 \times 10^{11} \mathrm{cm}^{-2}$ (upward band bending), $s_{\mathrm{GB,0}}^n$ to the interval 100-1000 cm/s. The blue bars stand for downward band bending, the red ones for upward band bending.}
 \label{fig:sGB_eval}
\end{figure}

Fig.~\ref{fig:sGB_eval} provides the simulations of experimental $s_{\mathrm{GB}}$ values, ranging from about 20 to 5200 cm/s, which were determined by means of CL on a high-efficiency Cu(In,Ga)Se$_2$ layer \cite{quirk_2024,krause_2020} with $N_{\mathrm{A}} = 2 \times 10^{16} \hspace{0.1cm} \mathrm{cm}^{-3}$.  We would like to remind that we assume similar (but not identical) defect properties at the corresponding grain boundaries and therefore, determine the value ranges for $N_{\mathrm{GB,charge}}$ and $s_{\mathrm{GB,0}}^n$ that can be used to simulate successfully the experimental $s_{\mathrm{GB}}$ values; i.e., these $N_{\mathrm{GB,charge}}$ and $s_{\mathrm{GB,0}}^n$ value ranges are always the same for each $s_{\mathrm{GB}}$ dataset.\\
 
It is apparent from  Figs.~\ref{fig:sGB_sGB0_NGB}a and b that experimental $s_{\mathrm{GB}}$ values ranging over several orders of magnitude (i.e., 20 to 5200 cm/s) cannot be simulated assuming small  $N_{\mathrm{GB,charge}}$ values on the order of $10^{9}$-$10^{10} \mathrm{cm}^{-2}$. Therefore, the simulations shown in Fig.~\ref{fig:sGB_eval} are based on $N_{\mathrm{GB,charge}}$ values restricted to $0.3$-$3$ (downward) and $0.9$-$9 \times 10^{11} \mathrm{cm}^{-2}$ (upward band bending). For $s_{\mathrm{GB,0}}^n$, a value range of 100-1000 cm/s, on the same order of magnitude as the median of the experimental $s_{\mathrm{GB}}$ values, 300 cm/s, was found to be appropriate. These intervals are highlighted in Figs.~\ref{fig:sGB_sGB0_NGB}a and b by a red and a yellow rectangle. \\

The ranges of $N_{\mathrm{GB,charge}}$ and $s_{\mathrm{GB,0}}^n$ that result in the corresponding, experimental $s_{\mathrm{GB}}$ values are given in Figs. \ref{fig:sGB_eval}a and b. Figs. \ref{fig:sGB_eval}c depicts the ranges for the band-bending values $\Phi_{\mathrm{GB}}$. It is apparent from Fig.~\ref{fig:sGB_eval} that several $s_{\mathrm{GB}}$ values can be modeled assuming both, downward and upward band bending. The true excess charge state of these grain boundaries cannot be revealed by the approach in the present work, but may be assessed by electrical analyses in scanning-probe microscopy, such as by conductive atomic force microscopy. Nevertheless, the ranges of $N_{\mathrm{GB,charge}}$ and $s_{\mathrm{GB,0}}^n$ (Fig.~\ref{fig:sGB_eval}) can be used as input parameters in multidimensional device simulations.\\



\newpage
In the Appendix Sections  \ref{sec:app_case_cigs} and \ref{sec:app_case_si}, evaluations of the $s_{\mathrm{GB}}$ values as the one shown in Fig. \ref{fig:sGB_eval} are demonstrated for (Ag,Cu)(In,Ga)Se$_2$ and Si solar-cell absorbers. It is apparent that basically, the $s_{\mathrm{GB}}$ values for an individual absorber material always exhibit distributions over several orders of magnitude and thus, the magnitudes as well as the distributions of $N_{\mathrm{GB,charge}}$, $s_{\mathrm{GB,0}}^n$, and $\Phi_{\mathrm{GB}}$ are similar to those in Figs. \ref{fig:sGB_eval}. 
Overall, we simulated the experimental $s_{\mathrm{GB}}$ values from various multicrystalline Si, (Ag,Cu)(In,Ga)Se$_2$, CdTe, kesterite-type, and halide-perovskite absorbers successfully by always using $N_{\mathrm{GB,charge}}$ and $s_{\mathrm{GB,0}}^n$ ranges on the same orders of magnitude for a specific absorber (not all data shown in the present work). Thus, the presented approach is a general procedure to estimate the orders of magnitude of the effective defect density and its capture cross-section for any photovoltaic absorber material.


\section{Correlation of excess charges and compositional changes at grain-boundary planes}

\begin{figure}[h!]
  \centering
  \includegraphics[width=10cm]{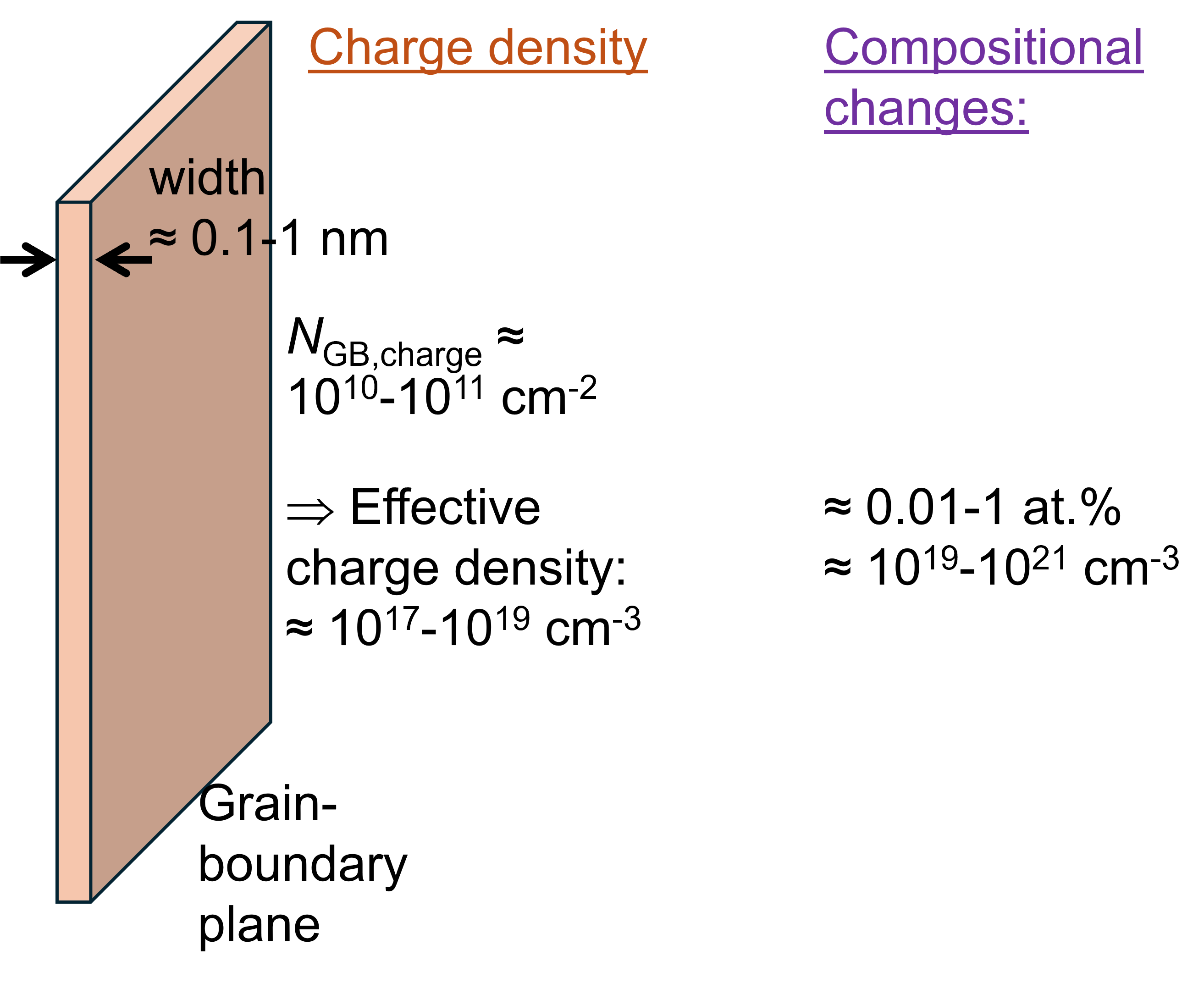}
\caption[]{Schematics of the charge density at a grain-boundary plane (left) and the magnitude of the compositional changes around this planar defect (right). From the considerations and results in the present work, the effective charge density can be assumed to be about $10^{17}$-$10^{19} \hspace{0.1cm} \mathrm{cm}^{-3}$, while the compositional changes typically detected via various microscopic analysis tools resides on larger orders of magnitude ($10^{19}$-$10^{21} \hspace{0.1cm} \mathrm{cm}^{-3}$); thus, effects of the local composition on the charge density present at a specific grain boundary can not be verified easily by the available analysis tools.}
 \label{fig:GB_charges+compo}
\end{figure}

In polycrystalline compound semiconductors, grain boundaries feature changes in composition within a very narrow range on the order of 0.1-1 nm. These changes have been interpreted as atomic / ionic reconstructions of the atomic planes adjacent to the grain boundary \cite{abou-ras_prl_2012,abou-ras_aem_2012}. It is a valid question of whether different $N_{\mathrm{GB,charge}}$ and $\sigma_n^{\mathrm{eff}}$ values, leading to substantiallly different recombination velocities $s_{\mathrm{GB}}$, are related to different types of compositional changes found at different grain boundaries \cite{abou-ras_pssrrl_2016,cojocaru-miredin_2018}. In order to address this question, it makes sense to regard the orders of magnitude of the charge densities at the grain-boundary plane (that are screened by the free charge carriers) as well as those of the detected compositional changes. Since the $N_{\mathrm{GB,charge}}$ values were found to be on the order of  $10^{10}$ to $10^{11} \hspace{0.1cm} \mathrm{cm}^{-2}$, and since the width of the grain boundary can be estimated to about 0.1-1 nm, the total charge density is about $10^{17}$-$10^{19} \hspace{0.1cm} \mathrm{cm}^{-3}$. On the other hand side, the changes in composition, also including impurity atoms, are typically on the order of 0.01-1 at.\%, i.e., about $10^{19}$-$10^{21} \hspace{0.1cm} \mathrm{cm}^{-3}$ (see also Figure \ref{fig:GB_charges+compo}). Thus, the change in the charge density $N_{\mathrm{GB,charge}}$ at a grain boundary can be orders of magnitude smaller than the changes in composition that can easily be detected by available characterization techniques with suitable spatial resolution and sensitivity (e.g., atom-probe tomography); under ideal conditions, the total charge density and the chemical sensitivity of the analysis tool (referring to the detection of lateral \textit{changes} in composition, not to the detection of impurities) may exhibit the same order of magnitude ($10^{19} \hspace{0.1cm} \mathrm{cm}^{-3}$). Nevertheless, in general, it is very difficult (if not impossible) to verify whether or not a direct correlation exists between changes in the charge density and changes in the composition.

\newpage

\section{Conclusions}
The present work revisited the determination procedures and magnitudes of recombination velocities at grain boundaries. It was described that experimentally measured recombination velocities $s_{\mathrm{GB}}$ can be simulated using specific value pairs of the charged defect density $N_{\mathrm{GB,charge}}$ and of the prefactor $s_{\mathrm{GB,0}}^n$. For one specific (experimental) $s_{\mathrm{GB}}$ value, always intervals of appropriate $N_{\mathrm{GB,charge}}$ and $s_{\mathrm{GB,0}}^n$ values are found. However, for the experimentally measured $s_{\mathrm{GB}}$ values ranging from several $10^0$ to several $10^4$ cm/s, the  intervals for $N_{\mathrm{GB,charge}}$ exhibit about the same orders of magnitude, about $10^{10}$ to $10^{11} \hspace{0.1cm} \mathrm{cm}^{-2}$, with resulting values for the prefactor $s_{\mathrm{GB,0}}$ of about $10^2$ to $10^3$ cm/s.  Finally, it has been outlined that the compositional changes at the grain boundaries that may be linked to changes in charged defect densities leading to enhanced nonradiative recombination are so small that their detection is very difficult using the currently available analytical tools. 

\section*{Acknowledgements}
The authors are grateful to Ulrike Bloeck (HZB) for assistance with the cross-sectional specimen preparation for SEM. The present work was supported by the project “EFFCIS‐II” funded by the Federal Ministry for Economic Affairs and Climate Action (BMWK) under contract numbers 03EE1059B (HZB) and 03EE1059C (MLU).


\section*{Appendix}

\appendix

\renewcommand{\theequation}{A.\arabic{equation}}
\setcounter{equation}{0}

\renewcommand{\thefigure}{A\arabic{figure}}
\setcounter{figure}{0}

\section{Derivation of band bending values at grain boundaries}

\subsection{Negative excess charge at grain boundary}
 \label{sec:app_upward_bb}

We assume a grain-boundary plane with negligible thickness and negative excess charge density $N_{\mathrm{GB,charge}}^-$ at position $x$=0. At sufficiently high net-doping density in a $p$-type semiconductor, this excess charge will be screened by free holes ($x \geq 0$):

\begin{equation}
\label{eqn:rho(x)}
\rho(x) = q N_{\mathrm{A}} A \exp(- x/w) ,
\end{equation}
  
where $q$ is the charge of a hole, $w$ is the screening length (on both sides of the grain-boundary plane.  Assuming a symmetrical distribution of $\rho(x)$ around $x=0$, the following equation must hold: 
\begin{equation}
\label{eqn:ampl1}
\begin{aligned}
& \frac{1}{q} \int_{0}^{\infty} \rho(x) \mathrm{d}x = \frac{N_{\mathrm{GB,charge}}^-}{2}\\
\Leftrightarrow & \, \frac{1}{q} \int_{0}^{\infty} q N_{\mathrm{A}} A \hspace{0.1cm} \exp(- x/w) \, \mathrm{d}x = \frac{N_{\mathrm{GB,charge}}^-}{2} \\
\Leftrightarrow &  \, A = N_{\mathrm{GB,charge}}^- / (2wN_{\mathrm{A}})
\end{aligned}
\end{equation}  

The electrical field is calculated via

\begin{equation}
\label{eqn:F(x)}
F(x) = - q N_{\mathrm{A}} A w / (\epsilon_0 \epsilon_{\mathrm{r}}) \exp(-x/w) ,
\end{equation}  

Integration of $F(x)$ results in the electrostatic potential: 

\begin{equation}
\label{eqn:phi(x)}
\phi(x) = - \frac{q N_{\mathrm{A}} A w^2}{2 \epsilon_0 \epsilon_{\mathrm{r}}} \exp(-x/w).
\end{equation}  

The hole density $p(x)$ 
\begin{equation}
\label{eqn:p(x)}
p(x) = \rho(x)/q + N_{\mathrm{A}} = N_{\mathrm{A}} A \exp(-x/w) + N_{\mathrm{A}} 
\end{equation}  

must fulfill the continuity equation

\begin{equation}
\label{eqn:conteq}
\begin{aligned}
& \mathrm{\overrightarrow{div}} \overrightarrow{j}(x) - \frac{\mathrm{d}}{\mathrm{d}t} p(x) = 0\\
& \Leftrightarrow \frac{\mathrm{d}}{\mathrm{d}x} j(x) = q \frac{\mathrm{d}}{\mathrm{d}x} \left[\mu p(x) F(x) - D \frac{\mathrm{d}}{\mathrm{d}x} p(x)\right] = 0 \\
& \Leftrightarrow p(x) F(x) - \frac{k_{\mathrm{B}}T}{q} \frac{\mathrm{d}}{\mathrm{d}x} p(x) = 0 \\
& \Leftrightarrow - p(x) \frac{\mathrm{d}}{\mathrm{d}x}\phi(x) - \frac{k_{\mathrm{B}}T}{q} \frac{\mathrm{d}}{\mathrm{d}x} p(x) = 0 
\end{aligned}
\end{equation}  
assuming a stationary case ($\frac{\mathrm{d}}{\mathrm{d}t} p(x) = 0$) and using Einstein's relationship $q D = \mu k_{\mathrm{B}}T$.

Applying Eqs. \ref{eqn:p(x)} and  \ref{eqn:phi(x)}, Eq. \ref{eqn:conteq} becomes
\begin{equation}
\label{eqn:conteq2}
\begin{aligned}
& \left[N_{\mathrm{A}} A \exp\left(-\frac{x}{w}\right) + N_{\mathrm{A}}\right] \frac{q N_{\mathrm{A}} A w}{\epsilon_0 \epsilon_{\mathrm{r}}} \exp\left(-\frac{x}{w}\right) - \frac{k_{\mathrm{B}}T}{q} \frac{N_{\mathrm{A}} A}{w} \exp\left(-\frac{x}{w}\right) = 0 \\
& \Leftrightarrow N_{\mathrm{A}} A \exp\left(-\frac{x}{w}\right) \left[\left(A \exp\left(-\frac{x}{w}\right) + 1 \right)  - \frac{k_{\mathrm{B}}T \epsilon_0 \epsilon_{\mathrm{r}}}{q^2 N_{\mathrm{A}} w^2} \right] = 0\\
& \Leftrightarrow \frac{N_{\mathrm{GB,charge}}^-}{2wN_{\mathrm{A}}} \exp\left(-\frac{x}{w}\right) + 1  - \frac{k_{\mathrm{B}}T \epsilon_0 \epsilon_{\mathrm{r}}}{q^2 N_{\mathrm{A}} w^2} = 0
\end{aligned}
\end{equation}
using $A = N_{\mathrm{GB,charge}}^- / (2wN_{\mathrm{A}})$ (Eq. \ref{eqn:ampl1}). In the following, we consider the limit case of large $N_{\mathrm{A}}$ and small $N_{\mathrm{GB,charge}}^-$, i.e., $N_{\mathrm{GB,charge}} / (2wN_{\mathrm{A}}) \exp\left(-x/w\right) \leq N_{\mathrm{GB,charge}}^- / (2wN_{\mathrm{A}}) \ll 1$; as outlined in the present work, this condition is fulfilled for various polycrystalline absorber materials. Thus, Eq. \ref{eqn:conteq2} becomes

\begin{equation}
\label{eqn:scrwidth}
\begin{aligned}
& 1  - \frac{k_{\mathrm{B}}T \epsilon_0 \epsilon_{\mathrm{r}}}{q^2 N_{\mathrm{A}} w^2}  = 0\\
&  \Leftrightarrow w^2  = \frac{k_{\mathrm{B}}T \epsilon_0 \epsilon_{\mathrm{r}}}{q^2 N_{\mathrm{A}}}\\
&  \Leftrightarrow w = \sqrt{\frac{k_{\mathrm{B}}T \epsilon_0 \epsilon_{\mathrm{r}}}{q^2 N_{\mathrm{A}}}},
\end{aligned}
\end{equation}
which is exactly the equation for the Debye screening length (see, e.g., Ref. \cite{Sze_2006}). For the amplitude $A$ follows
\begin{equation}
\label{eqn:ampl2}
\begin{aligned}
A = & \frac{N_{\mathrm{GB,charge}}^-}{2wN_{\mathrm{A}}}\\
= & \frac{q N_{\mathrm{GB,charge}}^-}{2 \sqrt{k_{\mathrm{B}}T \epsilon_0 \epsilon_{\mathrm{r}} N_{\mathrm{A}}}}.
\end{aligned}
\end{equation}
We can use the equations for $w$ and $A$ (Eqs. \ref{eqn:scrwidth} and \ref{eqn:ampl2}) to determine the electrostatic potential $\phi$ (Eq. \ref{eqn:phi(x)}) at position $x=0$:
\begin{equation}
\begin{aligned}
\label{eqn:phi(x)2}
\phi(x=0) = & - \frac{q N_{\mathrm{A}}}{2 \epsilon_0\epsilon_{\mathrm{r}}} \frac{q N_{\mathrm{GB,charge}}^-}{2 \sqrt{k_{\mathrm{B}}T \epsilon_0 \epsilon_{\mathrm{r}} N_{\mathrm{A}}}} \frac{k_{\mathrm{B}}T \epsilon_0 \epsilon_{\mathrm{r}}}{q^2 N_{\mathrm{A}}}\\
= & - \frac{\sqrt{k_{\mathrm{B}}T} N_{\mathrm{GB,charge}}^-}{4 \sqrt{\epsilon_0 \epsilon_{\mathrm{r}} N_{\mathrm{A}}}}.
\end{aligned}
\end{equation}  

The upward band bending $\Phi_{\mathrm{GB}}^{\mathrm{up}} = -q \phi(x=0)$ results to 

\begin{equation}
\label{eqn:Phi_GB_app}
\Phi_{\mathrm{GB}}^{\mathrm{up}} = \frac{\sqrt{k_{\mathrm{B}}T} q N_{\mathrm{GB,charge}}^-}{4 \sqrt{\epsilon_0 \epsilon_{\mathrm{r}} N_{\mathrm{A}}}}.
\end{equation}  

%
%
%

\subsection{Positive excess charge at grain boundary}
 \label{sec:app_downward_bb}

We assume a grain-boundary plane with negligible thickness and positive excess charge density $N_{\mathrm{GB,charge}}^+$ at position $x$=0. At sufficiently high net-doping density in a $p$-type semiconductor, this excess charge will be screened by ionized acceptors:

\begin{equation}
\label{eqn:rho_p(x)}
\rho(x) = - q N_{\mathrm{A}}.
\end{equation}

The electrical field is calculated via

\begin{equation}
\label{eqn:F_p(x)}
F(x) = \frac{q N_{\mathrm{A}} (w+x)}{\epsilon_0 \epsilon_{\mathrm{r}}} ,
\end{equation}  

Integration of $F(x)$ results in the electrostatic potential for $-w \leq x \leq w$:

\begin{equation}
\label{eqn:phi_p(x)}
\phi(x) =  \frac{q N_{\mathrm{A}} (w+x)^2}{2 \epsilon_0 \epsilon_{\mathrm{r}}}.
\end{equation}  

Using $N_{\mathrm{A}} = N_{\mathrm{GB,charge}}^+ / (2 w)$, the downward band bending $\Phi_{\mathrm{GB}}^{\mathrm{down}} = -q \phi(x=0)$ results in 

\begin{equation}
\label{eqn:Phi_GB_down}
\Phi_{\mathrm{GB}}^{\mathrm{down}} = - \frac{\left(q N_{\mathrm{GB,charge}}^+\right)^2}{8 \epsilon_0 \epsilon_{\mathrm{r}} N_{\mathrm{A}}}.
\end{equation}  

\newpage

%
%
%
%

\section{Case studies of recombination velocities at grain boundaries in (Ag,Cu)(In,Ga)Se$_2$ and microcrystalline Si solar-cell absorbers}

\subsection{Case study I: evaluation of recombination velocities at grain boundaries in (Ag,Cu)(In,Ga)Se$_2$ solar-cell absorbers}
 \label{sec:app_case_cigs}
 
 The following results were gathered on a Cu(In,Ga)Se$_2$ solar-cell absorber with [Ga]/([Ga]+[In]) (GGI) ratio of 0.34 and on a (Ag,Cu)(In,Ga)Se$_2$ solar-cell absorber with [Ag]/([Ag]+[Cu]) (AAC) ratio of 0.14. We note that the results shown below are not representative for (Ag,Cu)(In,Ga)Se$_2$ layers with similar GGI ratios.
 
%
\begin{figure}[!htbp]
  \centering
  \includegraphics[width=14.5cm]{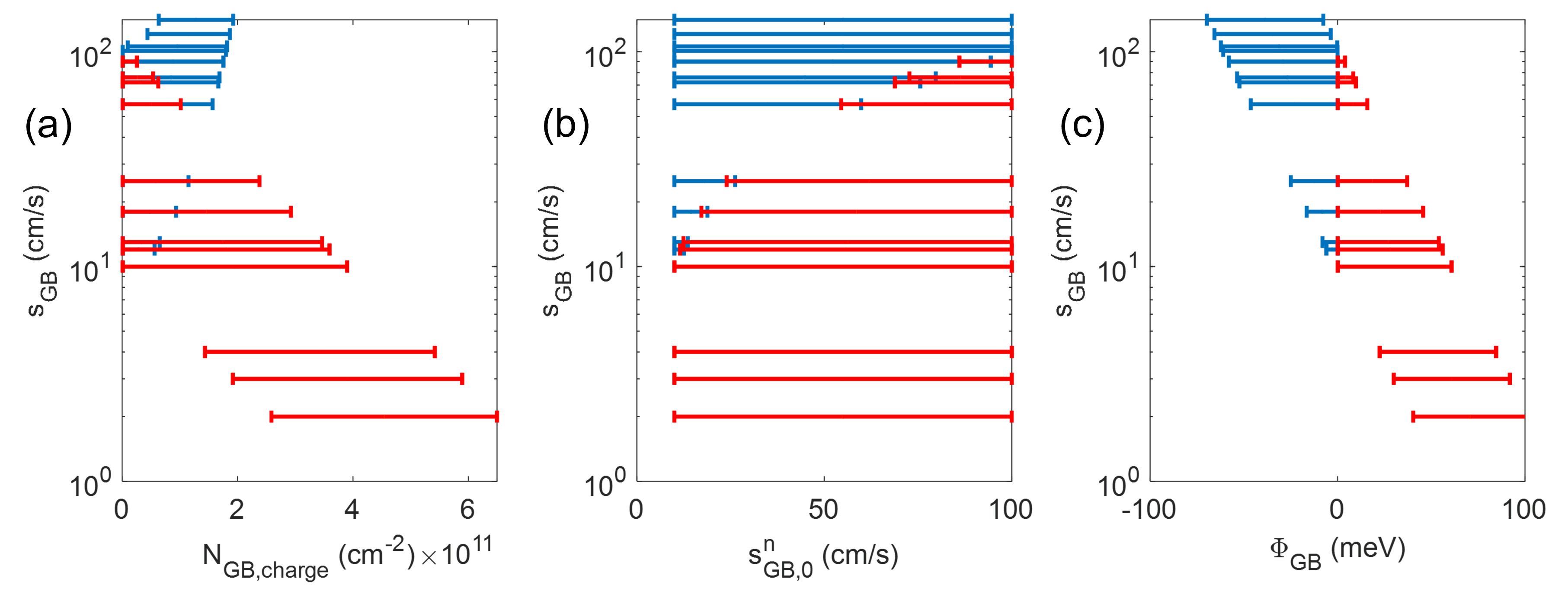}
\caption[]{Experimental $s_{\mathrm{GB}}$ values from a Cu(In,Ga)Se$_2$ layer with GGI=0.34 \cite{thomas_2024_pip_1} as a function of (a) the defect density $N_{\mathrm{GB,charge}}$, (b) of the prefactor $s_{\mathrm{GB,0}}^n$, as well as (c) of the band bending $\Phi_{\mathrm{GB}}$. The net-doping density was $N_{\mathrm{A}} = 1 \times 10^{16} \hspace{0.1cm} \mathrm{cm}^{-3}$, while $N_{\mathrm{GB,charge}}$ was restricted to $0.2$-$2$ (downward) and $0.7$-$7 \times 10^{11} \mathrm{cm}^{-2}$ (upward band bending), $s_{\mathrm{GB,0}}^n$ to the interval 10-100 cm/s (the median of the experimental $s_{\mathrm{GB}}$ values was about 30 cm/s). The blue bars stand for downward band bending, the red ones for upward band bending.}
\label{fig:sGB_eval_CIGSe_GGI_034}
\end{figure}
\begin{figure}[htbp!]
  \centering
  \includegraphics[width=14.5cm]{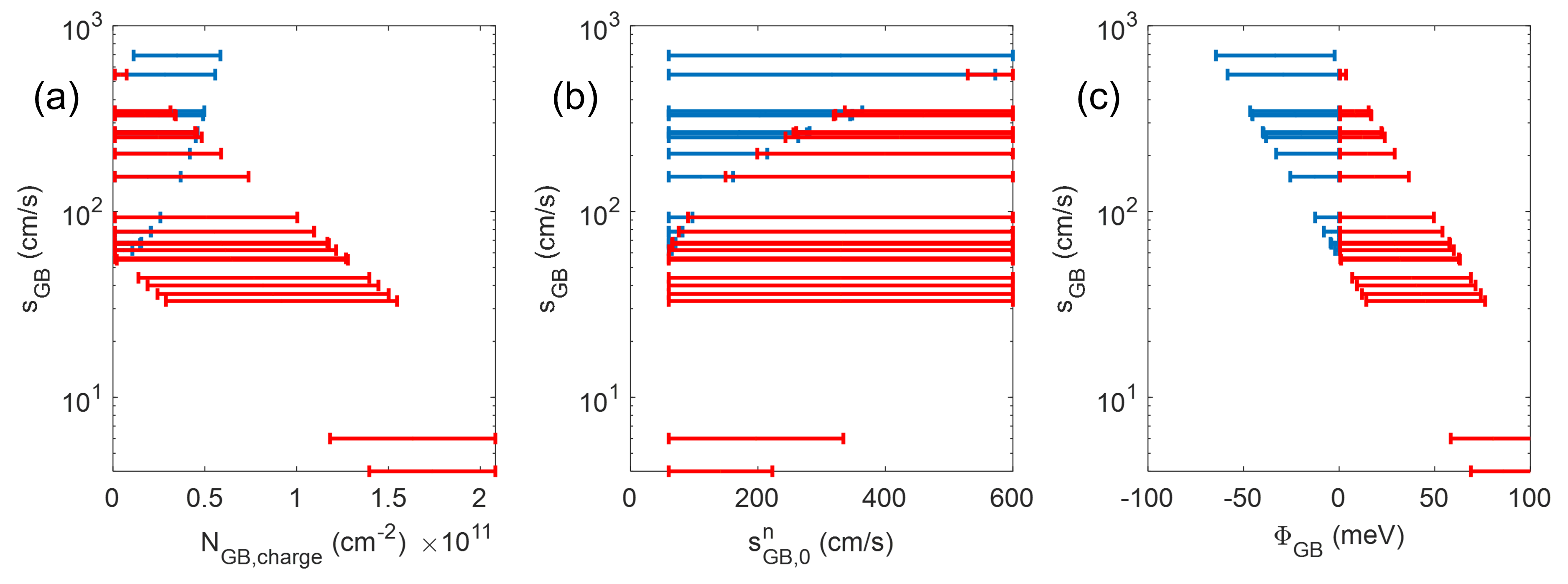}
\caption[]{$s_{\mathrm{GB}}$ values from a (Ag,Cu)(In,Ga)Se$_2$ layer with AAC=0.14 \cite{thomas_2024_pip_2} as a function of (a) the defect density $N_{\mathrm{GB,charge}}$, (b) of the prefactor $s_{\mathrm{GB,0}}^n$, as well as (c) of the band bending $\Phi_{\mathrm{GB}}$. The net-doping density was $N_{\mathrm{A}} = 1 \times 10^{15} \hspace{0.1cm} \mathrm{cm}^{-3}$, while $N_{\mathrm{GB,charge}}$ was restricted to $0.06$-$0.6$ (downward) and $0.2$-$2 \times 10^{11} \mathrm{cm}^{-2}$ (upward band bending), $s_{\mathrm{GB,0}}^n$ to the interval 60-600 cm/s (the median of the experimental $s_{\mathrm{GB}}$ values was about 100 cm/s). The blue bars stand for downward band bending, the red ones for upward band bending.}
\end{figure}

\newpage

\subsection{Case study II: evaluation of recombination velocities at grain boundaries in wafer-based, microcrystalline Si solar-cell absorber}
\label{sec:app_case_si}
 
The $s_{\mathrm{GB}}$ values in the following viewgraph were extracted from Sio et al. \cite{sio_2016}, who determined the recombination velocities at grain boundaries by means of photoluminescence imaging in multicrystalline Si wafers which underwent various treatments (gettered, hydrogenated, as well as gettered and hydrogenated). An interesting fact from this work is that the treated Si wafers do not exhibit smaller $s_{\mathrm{GB}}$ values than the as-cut Si wafer. It is noted by Sio et al. \cite{sio_2016} that the applied procedure cannot detect $s_{\mathrm{GB}}$ values of smaller than about 200 cm/s, which definitely has an impact on the magnitudes of the extracted $N_{\mathrm{GB,charge}}$ and $s_{\mathrm{GB,0}}^n$ values.
 
 \begin{figure}[htbp!]
  \centering
  \includegraphics[width=14.5cm]{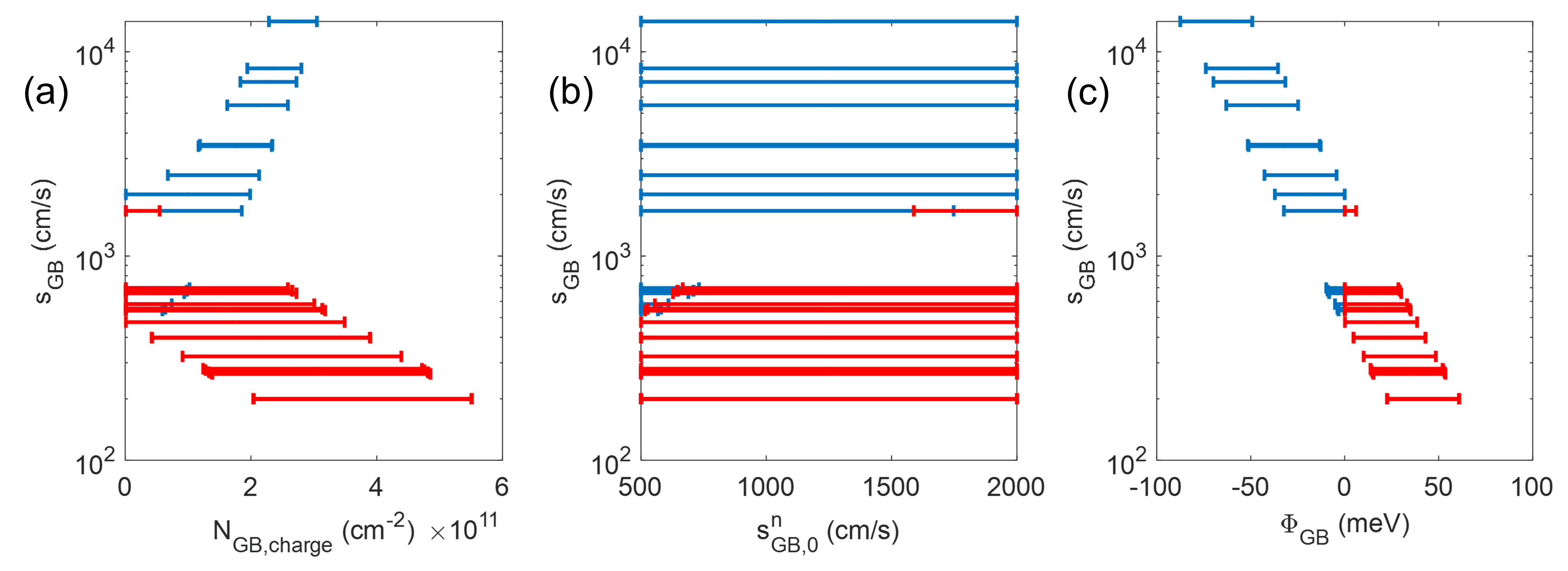}
\caption[]{$s_{\mathrm{GB}}$ values from a multicrystalline Si wafer (as-cut) \cite{sio_2016} as a function of (a) the defect density $N_{\mathrm{GB,charge}}$, (b) of the prefactor $s_{\mathrm{GB,0}}^n$, as well as (c) of the band bending $\Phi_{\mathrm{GB}}$. The net-doping density was $N_{\mathrm{A}} = 2 \times 10^{16} \hspace{0.1cm} \mathrm{cm}^{-3}$, while $N_{\mathrm{GB,charge}}$ was restricted to $0.2$-$2$ (downward) and $0.7$-$7 \times 10^{11} \mathrm{cm}^{-2}$ (upward band bending), $s_{\mathrm{GB,0}}^n$ to the interval 200-2000 cm/s (the median of the experimental $s_{\mathrm{GB}}$ values was about 1000 cm/s). The blue bars stand for downward band bending, the red ones for upward band bending.}
\end{figure}

\newpage

\bibliographystyle{pf}
\bibliography{GB_analyt_exp}

\end{document}